# From nuclear safety to LLM security: Applying non-probabilistic risk management strategies to build safe and secure LLM-powered systems

Alexander Gutfraind[1,2*] and Vicki Bier[3]

May 20, 2025

**Abstract**

Large language models (LLMs) offer unprecedented and growing capabilities, but also introduce complex safety and security challenges that resist conventional risk management. While conventional probabilistic risk analysis (PRA) requires exhaustive risk enumeration and quantification, the novelty and complexity of these systems make PRA impractical, particularly against adaptive adversaries. Previous research found that risk management in various fields of engineering such as nuclear or civil engineering is often solved by generic (i.e. field-agnostic) strategies such as event tree analysis or robust designs. Here we show how emerging risks in LLM-powered systems could be met with 100+ of these non-probabilistic strategies to risk management, including risks from adaptive adversaries. The strategies are divided into five categories and are mapped to LLM security (and AI safety more broadly). We also present an LLM-powered workflow for applying these strategies and other workflows suitable for solution architects. Overall, these strategies could contribute (despite some limitations) to security, safety and other dimensions of responsible AI.

## 1 Introduction

In light of the meteoric growth of the capabilities of LLMs in recent years, some observers have predicted that AI could be applied across the economy, and compared it to revolutionary inventions such as the internet or computing (Agrawal et al., 2022). However, such capabilities bring with them risks of adversarial penetration, misuse and accidents (Hacker et al., 2023). While classic machine-learning (ML) algorithms already suffer from insufficient training data or embedded biases (Kleinberg et al., 2018), the problem escalates with generative AI.

Current and proposed approaches to these risks include regulations (Hacker, 2023; Miotti and Wasil, 2023), well-defended software architecture (Mozes et al., 2023; Rae et al., 2024), and audits and controls (Bengio et al., 2023; Schuett, 2023). However, these approaches have already been breached in multiple incidents (Wei et al., 2023), provide few hard guarantees and precious little defense in depth, and are already being overmatched by rapidly advancing offensive AI technologies (Mehrotra et al., 2023).

Therefore, in this paper, we draw attention to promising approaches to risk management that could be drawn from engineering, medicine, and other fields. In particular, we single out *qualitative* strategies that are relatively easy to use, and robust to uncertainty about the event space of hazards and their probabilities. These strategies are useful for designers and software engineers who want to build LLM-powered solutions in a safe and responsible manner, as well as organizations that want to acquire LLM-powered solutions but are concerned about the risks. After reviewing the space of qualitative risk-reducing strategies, we show how they could be applied. Overall, we believe that they offer effective risk mitigation that complements existing strategies for building trustworthy LLM-powered systems.

The novel aspects of our work are the following: (1) introducing a risk-management framework for AI (focused on LLM-powered systems) that could be applied in the common situation of poor data on risks; and (2) introducing an arsenal of 100+ strategies developed in other fields that could be applied to AI, highlighting strategies that are rare or novel to software engineering.

## 2 Risk management frameworks

In the discipline of risk management, risk is usually defined through the tools of probability theory. In this formulation, the space of possible adverse events can be enumerated and denoted as $W$, and each event $i \in W$ is assigned an estimated frequency $f_i$ per year and consequence $c_i$ per event. The total risk per year is then the weighted sum:

$$R = \sum_{i \in W} f_i c_i$$

Probabilistic risk assessment (PRA) (Kumamoto and Henley, 1996) has been found to be useful in cases where the undesirable outcomes of interest are clearly identified and the biggest challenge is estimating their probabilities. Once the largest contributors to risk have been estimated, the decision of what to do about them may be straightforward, especially if relatively inexpensive risk-reduction options are available (Bier, 1996). Decision theoretic-approaches (Gilboa, 2010) can then examine alternative risk mitigations, trade-offs, nonlinearities, and dependencies (Keeney and Raiffa, 1993).

[1]Amazon Web Services and [2]Loyola University of Chicago. [3]University of Wisconsin at Madison. *Send correspondence to sgfriend@amazon.com

In theory it is possible to map any risk-management problem to the probabilistic framework (Gilboa, 2010). However, PRA is not easily applicable to open-ended problems outside Savage's small world setting (Savage, 1951), where the sample space of possible outcomes cannot be enumerated or the probabilities are difficult to quantify (Shafer, 1986). This situation, sometimes called Knightian or radical uncertainty (Kay and King 2020), occurs frequently in engineering, management, medicine, finance, military science, and many other disciplines. In response to these challenges, these fields have developed an arsenal of strategies for managing risk, some of which could be applied to building AI-powered systems.

Our proposal to manage risks in LLM-powered systems starts from the observation that there exist generic (i.e., cross-field) strategies for managing risk ((Todinov, 2015), and such strategies could be applied or adapted to novel fields. (Todinov, 2007) catalogued these strategies within fields of engineering, while (Gutfraind, 2023) also included strategies from medicine, management sciences, and other fields. The latter study published a catalogue of strategies, termed Risk-reducing design and operation toolkit (RDOT) that includes many counter-adversarial strategies from fields like game theory and political sciences. The RDOT catalogue has five categories of risk-reducing strategies (with some overlaps):

1. Structural strategies that design or improve preparedness for uncertainty
2. Reactive strategies that improve detection of events and subsequent response to them
3. Formal strategies involving algorithms or workflows
4. Strategies against adaptive adversaries
5. Strategies involving multi-stage or long-term risk management

**Structural strategies:** Briefly, these strategies involve designing and developing systems to be robust and resilient to uncertainties. Representative examples of this class of strategies are enhancing resilience (Ganin et al., 2016), spatial separation (Todinov, 2015), multi-layered defenses (Nunes-Vaz et al., 2011), and the use of damage-absorbing sacrificial parts (Short et al., 2018). There are also non-technical structural strategies, such as contingency planning, safety culture, planned delegation (or centralization) of authority, and regulation (either internal to an organization or by an outside body), which could for example mandate processes to reduce risk.

**Reactive strategies:** These strategies involve detection and interdiction of active threats. These include early-warning systems, techniques for detection, and methods for improving the response to detected incidents. Examples include interdiction at stand-off range, incident- response units (Ruefle et al., 2014), fault containment, and forensics and attribution.

**Formal methods:** These are strategies that use systematic decision-making processes or computational algorithms to discover and reduce risks. They include approaches such as event-tree analysis (Ferdous et al., 2009), system simulation (Henderson and Nelson, 2006), stochastic programming (Wallace and Ziemba, 2005), and others.

**Strategies for adaptive adversaries:** These are structural, reactive, or formal strategies that are designed specifically to protect against an adaptive adversary (Rios Insua et al., 2009). The adversary might be a business competitor, a political or military enemy, or an AI-like adaptive intelligence or algorithm. Examples of strategies to protect against adaptive adversaries include intentional misdirection through secrecy and/or deception (Zhuang and Bier, 2011), strategies for arms races, and others.

**Strategies for multi-stage or long-term planning under uncertainty:** These strategies involve improving sequences of decisions over time. Often, an investment at an early stage can enable adoption of more favorable options in subsequent stages, including some of the structural or counter-adversarial strategies mentioned earlier. Examples of such strategies include scientific research, sequential prototyping, and forms of hedging (Brach, 2003).

As evidence of the potential of RDOT, a recent study has applied it to the problem of reducing hallucinations in a LLM-powered system. This system was a previously deployed solution that utilized LLM and a knowledge base in a retrieval-augmented generation (RAG) architecture (Lewis et al., 2020). On reviewing the 100+ strategies in RDOT, 15% were judged as highly promising and not previously considered in this context. Another 31% were previously considered or already utilized in the project (Shi and Gutfraind, 2025). However, the study did not consider the potential of RDOT to improve security in adversarial settings.

## 3 Adapting the framework

Generalizing from a single project, in this section we demonstrate how strategies from each of the categories in RDOT could be applied to risks in LLM-powered systems, identifying relevant literature in each category. To demonstrate the power of these methods, we discuss well-known and successful examples from each category, many of which are familiar to software engineers: multi-layer defense; intrusion detection; progressive deployment; etc. We lack the space to consider individual applications to AI of the 100s of risk-reduction strategies, and therefore invite readers to explore the full catalog of such strategies and match them to their areas of application (see Table A1 in the Appendix). The most novel aspect of our proposal is to point out risk-management strategies coming from high-risk fields such as nuclear engineering, aeronautics, and others. Many of these are uncommon in AI or entirely novel to AI, including failure modes, effects, and criticality analysis (FMECA), bounding analysis, checkrides, and spiral development. As the field matures and AI enters higher-consequence applications, these might be quite useful.

### 3.1 AI risk reduction through structure and design

An overarching risk-mitigation principle in many fields of engineering may be termed robustness (Ezell et al., 2001; Bier and Gutfraind, 2017). For example, devices such as elevators, aircraft, and electric circuits are typically built with a factor of safety (Zheng et al., 2006), so that unexpected loads are unlikely to cause damage. Robustness also ensures that any errors in system design or modeling would be unlikely to jeopardize the system. Additionally, robustness allows system engineers to dispense with determining in advance exactly how a system will be used (i.e., the allowable ranges of inputs, procedures, and so on). Robustness can also be used in less tangible settings–e.g., in financial engineering (where financial planners keep reserves of cash or credit lines), or in software engineering (where programmers explicitly attempt to reduce dependencies between software components). There are a variety of approaches to achieving robustness–resilience (Ganin et al., 2016), fail-safe design, firebreaks, etc. These defenses are often combined to form an overall multi-echelon or multi-layer defense (Nunes-Vaz et al., 2011), in order to provide a more complete set of barriers and achieve overlapping coverage to decrease the risk of unexpected gaps.

AI systems could find use risk-reducing designs inspired by this cluster of strategies. For example, in creating training data for chatbots, a robust design (already being used) is to specify a safety margin regarding what training data could be used, in order to reduce the risks of copyright violations or poisoning. Similarly, in scanning inputs and outputs of LLMs, multiple heterogeneous filters might provide a multi-echelon defense against data exfiltration. The overall software could also be designed to be fail-safe; e.g., if an anomaly occurs, the system could be programmed to go into a safe state such as human intervention.

Beyond technical methods, organizational solutions often serve as an indispensable way of addressing risks, so we draw attention to the idea of safety culture. In areas such as nuclear engineering or clinical medicine, organizations use training and audits to build and maintain organizational practices for seeking out, detecting, and eliminating risks, and encourage questioning of potentially hazardous practices or circumstances (Gershon et al., 2004). Evidence suggests that insufficient attention is paid to the need for safety culture in software engineering, where safety may be seen as a hurdle to building powerful software quickly (Anderson, 2001). As AI becomes more powerful, safety culture may become more relevant, and play the type of central role that it does in the nuclear and aerospace industries.

Another important class of techniques for risk reduction is technical standards and regulation. These include requirements for third-party audits (say, for certain types of software tools, or use of such tools in critical application areas), and the establishment of best practices or professional standards, as was recently announced by US NIST (Tabassi, 2023). Some regulations could be binding at all times, while others could be binding only when observations suggest heightened risk. As suggested by Marcus (Marcus, 2024), the implementation could be at different levels–internally within companies, by national agencies (e.g., analogous to the U.S. Nuclear Regulatory Commission), and globally (analogous to the International Atomic Energy Agency).

### 3.2 Risk reduction through detection & interdiction

In many fields of safety engineering, waiting for hazards to materialize and then attempting to resolve them after the fact may be too costly, slow, or dangerous (Elbakidze and McCarl, 2006). In those contexts, risk reduction is best achieved by detecting a hazardous event early, and either interdicting it in real time, or implementing preplanned near-immediate responses. Examples include use of remote sensing to detect an approaching hazard (such as a storm or an incoming missile), and close monitoring of a manufacturing system to detect an anomalous condition (e.g., equipment vibration) before it causes a failure. Detection is often coupled to interdiction; i.e., activities that nullify or limit the damage through methods such as interception at a distance, isolation, or dissipation (Sadiq and McCreight, 2013).

### 3.3 Formal methods for proactive risk discovery & reduction

Here, formal methods refer to procedures, workflows, and algorithms that can help to identify risks before a system is put into operation (Leveson, 2012). While some such methods are quantitative (e.g., optimization, system simulation), we focus here on qualitative methods such as fault-tree analysis, checklists, and other approaches that can be applied to smaller projects without high levels of mathematical knowledge on the part of practitioners.

The paradigmatic method for risk identification is Scenario analysis; see for example (French, 2022; French, 2023) and references contained therein. Scenario analysis, even if not comprehensive, can give system designers and other decision makers a feel for the range of possible outcomes from a given design decision or hazard. While commonly used as part of PRA, scenario analysis can also serve as a qualitative method of problem structuring and scenario identification with no attempt at quantification. Other qualitative risk methods are also available, such as hazard and operability studies (HAZOPS), which is widely used in the process industries (Redmill et al., 1999), and Failure modes, effects, and criticality analysis (FMECA) ((Bowles, 1998). HAZOPS is an inductive elicitation process, prompting analysts to think of what could possibly go wrong. By contrast, FMECA is deductive in nature, beginning with a list of failure modes and characterizing their effects. At a higher level, (Leveson, 2012) has proposed a system-theoretic accident model and processes, which is specifically designed to identify potential risks in the face of the complexities, interactions, and nonlinearities that can be associated with software.

### 3.4 Advanced counter-adversarial strategies

Many of the best defenses against adversaries are structural methods inspired by good engineering and software-architecture practices (see Table A1 in the Appendix). Well-known examples are designs that minimize the attack surface for malicious hackers, or introduce gaps and firebreaks between users and sensitive data. Software architects often use data-dispersed storage or replication to prevent catastrophic losses due to a single failure, or losses due to ransomware. Reactive strategies, as discussed earlier, can also offer protections against adversaries, through the use of methods such as intrusion detection and honeypots. As LLMs become more capable, we must increasingly consider LLMs as potential adversaries and apply these strategies against them, protecting against both possible malfunctions of the LLM contained within the system being designed as well as possible LLM-powered intruders.

More advanced counter-adversarial strategies originating in fields such as military science and counter-terrorism (Bier and Azaiez, 2011; Rios Insua et al., 2009) could also be applied for securing LLMs. Other strategies are inspired directly or indirectly by game theory; e.g., minimax, randomization (Jain et al., 2010), and others. Applying game theory requires fairly sophisticated analysis, but can offer guaranteed security even against highly-resourced adversaries (see Table A1 in the Appendix).

### 3.5 Multi-stage and long-term planning for AI risk

Certain risk-mitigation strategies can be carried out only through multi-stage decision-making or long-term investments spanning months and years. Because we cannot forecast the future with any confidence, we instead try to improve our ability to respond in the future (Goble et al., 2018).

A basic multi-stage tool for organizations adopting AI is knowledge dissemination–equipping developers and users with knowledge about AI risks and strategies to effectively reduce them. This will allow to upskill developers and users of AI who are new to the pitfalls and hazards of the technology they are building or using. Another strategy is sequential prototyping–a technique often used in long-term product development to understand a system and its potential risks. While the software-development community understands the value of iterative development (Ries, 2011), the approach assumes that the system is regularly released into the hands of users. By contrast, as AI technologies become more powerful and potentially harmful, there may be an increased need to carefully plan and control the prototyping and release processes to identify and mitigate risks before the software is released. Such processes are already used in the robotics community (Afzal et al., 2020).

### 3.6 Selecting strategies for specific problems

How could one begin adopting some of these risk-management strategies in a specific AI system? In our experience building AI solutions, the starting point is to evaluate the project along the following dimensions: (1) the impact (i.e., the value and complexity of the system, or the consequences of failure); (2) the available resources for the project; and (3) whether adversarial threats are relevant. These dimensions can rapidly filter down the number of strategies that need to be considered, from the 100+ available to a few dozen.

In the case of low-complexity and low-value systems, the most suitable mitigation strategies tend to be structural and passive. Examples of such low-complexity, low-value systems are tools for data reporting that have low business impact, do not alter data, and are used by trained internal users, rather than by customers. The risks of such systems can typically be adequately controlled by following generic good engineering-design practices such as fail-safe, modular design, access control, and so forth. Reactive strategies can be suitable for projects when it is relatively inexpensive to build the necessary reactive logic, for example in pure software systems and when there is sufficient time to intervene after detection of a problem.

Projects of higher complexity or system value should consider additional categories of strategies. Complex software systems, including AI software, are often embedded in large organizations, and therefore risk management should include assisting these organizations in the safe use of AI. This assistance can include structural strategies such as training, incident-response units, and improved command and control. It may also be cost-effective to invest in reactive risk-management strategies such as automatic anomaly detection and containment. Lastly, it will likely be cost-effective to implement formal strategies such as hazard identification and post-release monitoring of the software. Many AI systems are potentially exposed to intentional cyber threats or other adversaries (e.g., malicious insiders, business competitors, out-of-control AI); these situations would suggest employing some specialized counter-adversarial strategies.

We have also had success using AI-powered workflow for matching risk-mitigation strategies to projects; for other uses of AI for risk analysis, see (Stødle et al., 2024)). In this workflow, we ask an LLM with reasoning capabilities (e.g., DeepSeek-R1; (Guo et al., 2025)) to find strategies that might be relevant to a given project, and report on how each component of the project could be secured from hazards. This is particularly helpful for finding strategies that are less familiar to the software engineer. To implement this approach, the LLM prompt needs two documents: (a) the list of strategies with descriptions; and (b) a detailed description of the project, including the components and their linkages. The LLM is then prompted to report strategies that could reduce the risk of each component using the available solutions. The raw output is usually infeasible or too complex, but contains some interesting suggestions. The engineer can then use experience to holistically evaluate possible designs and select a risk-management solution that would be feasible and efficient.

## 4 Conclusions

We have demonstrated here the application of risk-management strategies from engineering and other fields to the problem of building safe and reliable LLM-powered systems. We show how five classes of strategies (structural, reactive, formal, counter-adversarial, and multi-stage strategies) could be applied to this problem. These strategies overcome some of the challenges involved in applying quantitative methods of risk analysis to AI. We argue that they represent pragmatic and cost-effective strategies to help managers and engineers build and deploy safer AI systems.

## 5 Limitations

Many non-probabilistic strategies are qualitative, and are best used in novel applications where data and experience is scarce. As more data becomes available, risk management can estimate frequencies of events and shift towards more data-informed probabilistic approaches, which may offer more cost-effective risk mitigations. Unfortunately, they are rarely available or practicable in current AI applications, as pointed out earlier. One challenge when applying non-probabilistic strategies is how to combine multiple strategies efficiently, because multiple overlapping safety systems can create cost and complexity, and even introduce new modes of failure, as observed in the Three Mile Island nuclear accident (Perrow, 2011). To address this limitation, engineers often take a prototype system through multiple iterations in which risk-control strategies are progressively refined, achieving streamlined complementary coverage of risks.

### Acknowledgments

The authors thank Christos Christodoulopoulos for feedback. Supported in part by NIH grant R01-AI158666.

# Appendix A

**Table A1: An abbreviated version of RDOT catalog**. The full dataset (contains examples, references and clarifying notes.

| Strategy for uncertainty | Definition | Category |
|---|---|---|
| Accelerate adaptation | Accelerate adaptation to the adversary by e.g. analyzing past moves of an adversary and developing new tactics | adversarial |
| Adjust planning horizon | Lengthen or shorten the planning horizon to match the ability to forecast | config - org |
| Amass resources | Make the arsenal of resources as large as possible | formal - alg |
| Anomaly detection and investigation | seek out anomalous readings and trigger alerts | reactive - ID |
| Asymmetric offsets | Invest in capability that can neutralize an opponent's possible strength | adversarial |
| Automatic containment system | A physical system that automatically responds to impacting events by containing the damage | reactive - respond |
| Basic research | Conduct research into the phenomenon trying to better understand its sources and dynamics | enabler |
| Blue/green deployment | use a mixture of new and old solutions in order to provide a fallback if the new solutions fail | config - system |
| Bluff | When playing against a risk-averse opponent, bluff | adversarial; harnessing |
| Canary detection | Use a proxy to detect hazards or bound the potential effect of hazards | reactive - ID |
| Conflict stabilization | Prevent a conflict from escaling to higher or less predictible levels | adversarial |
| Consolidate components | Reduce the number of independent agents or degrees of freedom to make it more predictable or manageable | config - org/system |
| Contest and appeal decisions | The ability of users to challenge the output with a third party, thus prevent gross errors | formal - alg; config - org |
| Contingency planning | Establishing thorough plans, procedures, and technical measures that can enable a system to be recovered as quickly and effectively as possible following a service disruption | config - org |
| Coordinate action | Coordinate the response of multiple actors to improve its effectiveness | config - org |
| Decision template | Make decision following the steps of others or by following a decision algorithm | formal - workflow |
| Decoy | Minimize the damage from intelligent hazard by create false targets | config - system |
| Deflect | Move the site impacted by the hazard in order to reduce the damage | config - system |
| Delay | Delay the start time of the hazard in order to reduce its effect | config - system |
| Delegate control | Delegate authority to local agents or stakeholders to select among alternatives within an overall mission | config - org |
| Denial strategy | Denying options to an adversary to reduce uncertainty about their action | adversarial |
| Diagnosis of exclusion | Establish the root cause of a pathology or anomaly by excluding all the alternatives | format - workflow |
| Dispersed storage | Disperse a critical resource to multiple sites in order to prevent a single event from causing total loss | config - system; adversarial |
| Early warning system | Predict and detect possible hazardous events and help minimize their devastating impact | reactive - ID |
| Eliminate input variables | Reduce uncertainty about a plan or system but eliminating variables affecting its operation or reducing their variability | config - system |
| Event forensics and attribution | Determine the type of an event and reduce uncertainty about its perpetrators, nature and consequences | reactive - ID; adversarial |
| Event tree analysis | Find risks by exploring events and their consequences to the system | formal - workflow |
| Evolvable design | An approach to design and development that focuses on creating systems that can be easily changed and adapted over time | config - system |
| Expansive analysis | Perform detailed analysis of options involving data collection, modeling and other means | formal - workflow |
| Expert elicitation and judgment | Convene experts in the phenomenon and ask for their recommendation | formal - workflow |
| Factor of safety | Design the system to be stronger than strictly necessary in order to make it survive unexpected events | config - system |
| Fail-safe design | a design that in the event of a fault fails in a safe way or degrades gracefully | config - system |
| Failure mode and effects analysis | Failure modes and effects analysis (FMEA) is a bottom-up (inductive) analysis approach to evaluate all the component failure and predict the impact of these failures | formal - workflow |
| Fault tree analysis | A formal approach for resolving the basic causes of a given undesired event | formal - workflow |
| Find good enough solution | Explore and reject actions until a solution is found that satisfies all constraints | formal - alg |
| Firebreak/compartalized design | Compartmentalize an at-risk system to prevent damage from spreading, or create firebreaks in an existing system | config - system |
| Game-theoretic analysis | Use game theory to analyze actions and risk of actions of other players | adversarial |
| Gather data | Collect data about the system, setting or environment to enable better decisions or quantitative strategies | enabler |
| Grow the funnel | Increase the number of favorable outcomes by increasing the number of trials | harnessing |
| Hazard and operability study | A hazard and operability study (HAZOP) is a structured and systematic examination of a planned or existing process or operation in order to find and evaluate problems that may represent risks to personnel or equipment, or prevent efficient operation | formal - workflow |
| Heuristic solution | Apply a heuristic on-the-spot to select a solution ignoring uncertainty | formal - workflow |
| Hypothetico-deductive method | a scientific method that uses inductive reasoning to make predictions and then tests those predictions with experiments | formal - workflow |
| Incident investigation | Establish a unit or agency responsible for investigating incidents to drive preventative measures | formal - workflow |
| Incident response unit | Establish a team dedicated to responding rapidly and effectively to hazards | config - org; reactive - response |
| Increase system transparency | Share the design and current state of a system with users, operators and third parties to reduce their uncertainty | config - org/system |
| Incremental development | Rather than building the entire solution, construct the solution incrementally, adding new features to learn about risks | enabler |
| Independent auditors | Use of an independent organization to review safety and risks | formal - workflow |
| Independent certification | Require all technology and/or operators to be certified by an organization unrelated to their employer | formal - workflow |
| Index building | Develop an index (a random variable) to describe the uncertain quantity in order to enable securitization, anomaly tracking etc | enabler |
| Index-driven decisions | Using an index of the phenomenon, estimate measures of volatility to drive the response | formal - alg |
| Insurance & financial instruments | Use contracts or financial instruments to indemnify against possible losses | config - system |
| Kill switch | A device or process that quickly shuts down a system | config - org/system |
| Knowledge dissemination | Condense and communicate existing knowledge to all members of the group in order to make them more effective | enabler |
| Lifeboat subsystem | Implement a subsystem that can protect human users from the disaster, or generally, prevent total loss | config - system |
| Long vega | Invest in assets that benefit from increased volatility | harnessing |
| Managerial assumption | Manager or commander reduces uncertainty by imposing limits or assumptions on his unit | formal - workflow |
| Maximization of expected utility | Select between actions using estimated probability of events and their outcomes | formal - alg |
| Mechanistic modeling | a type of mathematical modeling that uses known physical principles to describe the behavior of a system and forecast its state | formal - alg |
| Meta-learning of unknowns | Identify gaps in knowledge in order to prioritize their closure or ensure safe degradation | enabler |
| Minimize area or time of impact | Shape the impacting event or the threatened system to minimize the area or time of impact | config - system |
| Misdirection/Deception | Cause the adversary to make incorrect action by creating a false impression of own action | adversarial |
| Modular design | Use independent modules in order to make the system easier to build and repair | config - system |
| Multi-layer defense | Design a system to absorb or respond to threats using multiple layers of defensive barriers | config - system |
| Operator checkrides | Require all new operators to be evaluated by an experienced co-pilot | formal - workflow |
| Opportunity discovery | an organizational unit or process whose role is collect and priority opportunities for gains | harnessing; config - org |
| Optimization of total outcome | Decide by maximizing the total welfare | formal - alg |
| Perimeter detection and response | Prevent damage from hazards by establishing a perimeter and detect threats as soon as they reach the perimeter | reactive - ID |
| Poll and aggregate | Collect data from experts and stakeholders and select action supported by a qualified majority, to reduce the risk of a mistake | formal - workflow |
| Portfolio rebalancing | Systematically adjust a portfolio of items to avoid excessive risk | formal - alg |
| Post-release monitoring | evaluate the system after its release to detect unexpected or dangerous behaviors | reactive - ID |
| Pre-release testing | evaluate the system before its deployment to detect unexpected or dangerous behaviors | formal - workflow; config - org |
| Precautionary principle (maximin) | Invest in preventing the worst-case outcome regardless of its probability | formal - workflow |
| Probabilistic risk management | Develop a probabilistic model of the risks and develop strategies to mitigate it efficiently | formal - alg |
| Prototype-driven development | Build a simple version of a complex solution in order to reduce risks and complexity, then progress to a more sophisticated version (or potentially a series thereof) | formal - workflow; enabler |
| Questioning attitude to anomalies | Staff are trained to question anomalies rather than merely check off boxes | config - org |
| Randomization of moves | Deliberately randomize actions in order to avoid being anticipated by an adversary | adversarial |
| Rapid identification after event | If unable to prevent adverse events, develop the capability to detect them rapidly in order to respond as quickly as possible and reduce ultimate damage | reactive - ID |
| Real options | Invest resources in obtaining a positive risk | harnessing; enabler |
| Red teaming | Discover risks by establishing a unit responsible for ethically attacking the system | formal - workflow; adversarial |
| Regulate by limiting | Set limitations on risky activities in order to reduce risk | config - org |
| Regulate by process | Establish processes and guidelines for doing risky activities in order to prevent accidents and other unwanted outcomes | config - org |
| Reinforcement learning | Iteratively try different actions (manually or algorithmically), measure the outcomes and optimize the actions | formal - alg |
| Resilient design | System designed to absorb, respond to, and recover from disasters and adapt to new conditions | config - system/org |
| Risk and control matrix | A technique for risk management that involves building a grid of undesirable events as rows and corresponding controls that could mitigate them | formal - workflow |
| Risk portfolio approach | Balance investments between different risks and opportunities to improve overall risk | formal - alg |
| rK strategy | Grow by either producing a large number of offsprings or few highly-capable ones | formal - alg |
| Robustness analysis | Evaluate a solution and its feasibility against uncertainty in parameters of the problem | formal - workflow |
| Sacrificial part | Design where functions and flows that are critical to system operation are protected through the sacrifice of less critical functions and flow exports | config - system |
| Safety culture | Create an organization culture that advances hazard prevention, investigation and mitigation | config - org |
| Sandboxing / Firewalling | Limit the operation of system to narrow domain of environments, users or inputs to mitigate unexpected behaviors or dangerous inputs | config - system |
| Self-disablement | Disarming mechanism - programmed self-disablement, self-disarmament or self-destruction | config - system |
| Simplify | Use simple solutions, designs and plans | formal - workflow |
| Stand-off response system | System that interdicts impacting events before their arrival | reactive - respond |
| Standard operating procedures | Define standard operating procedures or plans for routine and exceptional situations | config - org |
| Statistical modeling | Develop models of past data (e.g. average and outlier) in order to estimate the likelihood and risk of hazardous events | formal - alg |
| Statistical process control | Apply statistical modeling to monitor and control the quality of a production process | formal - alg |
| Stochastic search | Optimize system design or operation by exploring different solutions particularly when preparing to uncertainty | formal - alg |
| Strategy enablement | Enable new strategies in future decision points | enabler |
| Strengthen weakest link | Reduce the risk of failure by finding and strengthening the weakest component | config - system |
| Swim lanes | Divide responsibilities in clearly defined roles to reduce uncertainty about areas of responsibility and accountability | config - org |
| System modeling (mathematical-predictive-computational) | Using a model develop a description of the system its environment in order to improve its design | formal - alg |
| Take another swing | Improve the odds of favorable outcomes by repeating trials | harnessing |
| Tighten component tolerances | Increase the quality requirements of system components in order to reduce risk of failures | config - system |
| Trap risktaker | When playing against a risk-taking opponent, set up traps | adversarial; harnessing |
| Trial-and-error | At attempt different actions with little or no guidance until a satisfactory action is found | formal - workflow |
| Unicorn hunting | in selecting lotteries (i.e. bets), seek lotteries that sometimes give extreme positive returns even when the typical pay has low value | harnessing |
| Use the default action | Respond to unexpected events by invoking the default action or standard operating procedure with minimal analysis of alternatives | formal - workflow |
| User screening and training | Screen and train operators of a system in order to reduce risks from system operations | config - org |
| Wait and see | Do nothing until the situation changes or new information comes to light | formal - workflow |
| Wait for opportunities | Delay action until the timing is more favorable | harnessing |

See full dataset online at https://zenodo.org/records/14277084